\documentclass[aps,preprint,showpacs]{revtex4}
\usepackage{amsfonts}
\usepackage{amsmath}
\usepackage{amssymb}
\usepackage{graphicx}
\usepackage{bm}

\begin{document}

\title{Spin-dependent magnetotransport through a ring\\ in the presence of  spin-orbit
interaction}
\author{X. F. Wang}
\email{xf_wang1969@yahoo.com}
\author{P. Vasilopoulos}
\email{takis@alcor.concordia.ca}
\affiliation{Department of Physics, Concordia University\\
1455 de Maisonneuve  Ouest, Montr\'{e}al, Qu\'{e}bec, Canada, H3G 1M8}


\begin{abstract}
The Schr\"{o}dinger equation for an electron in a mesoscopic ring,
in the presence of the Rashba and
linear Dresselhaus terms of the spin-orbit interaction (SOI) and of a  magnetic field $B$, is solved exactly.  The effective electric fields of these terms as well as $B$  have
perpendicular and radial components. The interplay between them and $B$
and their influence on the spectrum is studied. The transmission through such a
ring, with two leads connected to it, is evaluated  as a function  of the SOI strengths and of the orientations
of these fields.
The  Rashba and
Dresselhaus terms affect the transmission in different ways.
The transmission through  a series of rings with different radii
and with SOI in both  arms of the rings or only in one of them is also evaluated.
For weak magnetic fields $B\leq 1$ T the influence of the Zeeman term on the transmission,
assessed  by perturbation theory, is negligible.
\end{abstract}

\maketitle

\section{Introduction}

The concept of the geometric phase \cite{berr,geom,ahar1,qian}
has attracted considerable interest
since it is established in a general way. In mesoscopic rings, the geometric phase
of electrons in magnetic and electric fields can be obtained by solving the time-independent
Schr\"{o}dinger equation. \cite{qian} From the point of view of the stationary
Schr\"{o}dinger equation, the energy
dispersion relation of electrons changes as the configuration of the system varies.
As a result, electrons
of the same energy have different wavevectors in different systems and may accumulate different
phases after passing even the same path in  real space. In the presence of
 external magnetic and electric fields, the one-particle Hamiltonian can be expressed as
\cite{sun}
\begin{equation}
H=(\bm{p}-e\bm{A}-\mu_B \bm{\sigma}\times \bm{E}/2c^2)^2/2m
\end{equation}
The contribution from the vector potential $\bm{A}$ corresponds to the Aharonov-Bohm (AB) phase \cite{ahar2}
and the contribution from
the
SOI to the Aharonov-Casher (AC) phase. \cite{ahar3}
In  mesoscopic systems of semiconductor heterostructures, however, the macroscopic SOI
results from the asymmetry of the microscopic crystal field
and may appear in different forms depending on the
materials and structures involved.
In materials with asymmetric crystal structure,
the cubic Dresselhaus SOI  term exists in
bulk materials while an extra linear Dresselhaus SOI (DSOI) term
appears in confined, low-dimensional systems due to the change of the crystal structure along the direction of the confinement. In systems with asymmetric confinement,
the Rashba SOI (RSOI) term results from a  non-vanishing
confining electric field as well as  from various other SOI mechanisms such as the one related to differing band discontinuities at the heterostructure interfaces considered in  {\bf k.p} models. The aggregate strength of all these SOI mechanisms is denoted by $\alpha$ \cite{feve}.

By inserting a mescoscopic ring into a circuit, we can study the quantum transport through the ring when the inelastic diffusion length is
larger than the size of the ring. In the two-terminal configuration,
the transmission depends on the
interference between electrons propagating through the ring's two arms and the transmission
properties through the two junctions connecting the ring and the leads. In general,
a symmetric junction can be described by a $3\times 3$ scattering matrix with the
transmission through each arm as a parameter. \cite{gefe,aron,yi}
For a ballistic, one-dimensional (1D) ring connected to two leads,
the scattering matrix can be determined by imposing the continuity of
the wave function and of the spin flux at each junction. \cite{xia,moln}
In the presence of SOI the transmission through the ring as well as the
geometric phases are spin-dependent
and the ring can be used as a spin-interference device \cite{nitt}.
Similar considerations apply to a square loop or arrays of such loops  \cite{kog}.
Theoretically this  spin interference was further  studied  in 1D and 2D rings
\cite{frus} but  only in the presence of the RSOI.

In this paper we study  ballistic  transport through one or more 1D rings,
symmetrically connected to two leads, in the presence of a magnetic field,
with components along the radial and perpendicular direction,
and of {\it both} terms of the SOI,  RSOI and DSOI. The corresponding {\it effective} electric fields have perpendicular and radial components. For weak magnetic fields, $B\leq 1$ T, the influence of the Zeeman term is validly assessed by perturbation theory. In Sec. II we present  the
one-electron energy spectrum and formulate the corresponding transfer-matrix transmission problem. In Sec. III we present   numerical results for the transmission, through one or more rings, and in Sec. IV concluding remarks.

\section{One-electron problem}

\subsection{Hamiltonian}
We consider a one-dimensional ring, of radius $a$,  in the $(x$-$y)$ or $(r$-$\theta$)
plane and in a magnetic field with components
$B_z=B\cos\gamma_3$ and
$B_r=B\sin\gamma_3$.

For the vector potential we choose the gauge
$\textbf{A}=(A_r,A_\theta,A_z)=(0,B_zr/2-B_rz,0)$ with $z=0$ in the plane of the ring.
The one-electron Hamiltonian is
\cite{loss}
\begin{equation}
H=(\hbar^2/2m^\ast a^2)(-i\partial/\partial\theta+\Phi/\Phi_0)^2+g\mu_B
\bm{\sigma} \cdot {\bf B}/2,
\end{equation}
where $\bm{\sigma}=(\sigma_x,\sigma_y,\sigma_z) \equiv (\sigma_r,\sigma_\theta,\sigma_z)$
are the Pauli matrices, $\Phi=B_z\pi a^2$ the magnetic flux passing through the ring,
$\Phi_0=h/e$ the flux quantum, $\mu_B$ the Bohr magneton, and $g$  the $g$
factor.
For a ring fabricated out of
a heterostructure the SOI can result
from asymmetric confinement along the $z$ direction
(RSOI, $H_{\alpha}$) or from the crystal structure changing along
the $z$ direction (DSOI, $H_\beta$). Considering both terms, we have
\cite{bych,dres,ross, sch}
\begin{equation}
H_\alpha =\frac{\alpha}{\hbar}\sigma_x(\hat{p}_y+eA_y)
-\frac{\alpha }{\hbar}\sigma_y(\hat{p}_x+eA_x),
\end{equation}
\begin{equation}
H_\beta =\frac{\beta}{\hbar}\sigma_x(\hat{p}_x+eA_x)
-\frac{\beta}{\hbar}\sigma_y(\hat{p}_y+eA_y).
\end{equation}
Here $\alpha\propto
\langle E_R \rangle$ and  $\beta\propto \langle E_D \rangle$
are the usual RSOI and  DSOI strengths, respectively,
and
$\langle\mathbf{E}_R\rangle=\langle E_R \rangle\mathbf{e}_z$ and $\langle\mathbf{E}_D\rangle=\langle E_D \rangle\mathbf{e}_z$, the corresponding
{\it effective} electric fields.
However, here we study the more general case in which  $\langle\mathbf{E}_R\rangle$ and $\langle\mathbf{E}_D\rangle$
can have a
radial component as well as has already been the case for the RSOI  \cite{qian,moro}.  Accordingly, we consider them
in the form
\begin{eqnarray}
\mathbf{E}_R&=&E_R(\sin\gamma_1\mathbf{e}_r+\cos\gamma_1\mathbf{e}_z),\\
\mathbf{E}_D&=&E_D(\sin\gamma_2\mathbf{e}_r+\cos\gamma_2\mathbf{e}_z).
\end{eqnarray}
That is, we assume the same form  of Eqs. (3) and (4) but with  the effective electric fields having  components along the  radial and  $z$ directions as specified in Eqs. (5) and (6).
Then the total Hamiltonian in the ring  reads
\cite{yi,meij}
\begin{eqnarray}
\nonumber
\hat{H}=\hbar \omega_0 [-i
\partial/\partial \theta+\phi+
\bar{\alpha}(\sigma_r\cos\gamma_1-\sigma_z\sin\gamma_1)
&+&\bar{\beta}\sigma_\theta(\sin\gamma_2-\cos\gamma_2)]^2\\*
+\hbar\omega_B(\sigma_r\sin\gamma_3+\sigma_z\cos\gamma_3),
\label{ham}
\end{eqnarray}
where $\phi=\Phi/\Phi_0$, $\omega_0=\hbar /2m^\ast a^2$,
$\omega_B=g\mu_B B/2\hbar$,
$\bar{\alpha}=\alpha a m^\ast/\hbar^2$,
and $\bar{\beta}=\beta a m^\ast/\hbar^2$.

In an isolated ring we can expand the wave function $\Psi$ in terms of an orthogonal and complete set
of eigenvectors $e^{in\theta}/\sqrt{2\pi}$ of the Hamiltonian $\hat{h}=-[\hbar^2/2m^\ast a^2]\partial^2/\partial^2\theta$
 with $n$ integer. The  expansion takes the form
\begin{equation}
\Psi=\frac{1}{\sqrt{2\pi}}\sum_n
\left(
\begin{array}{c}
C_n^+ e^{i n\theta}\\
C_n^- e^{i n\theta}
\end{array}
\right),
\end{equation}
where $C_n^+$ and $C_n^-$ are the coefficient of the spin-up and  spin-down eigenstates, respectively.  We can then write the secular equation $\hat{H}\Psi=E\Psi$ as
\begin{equation}
\left[
\begin{array}{cc}
\Omega_n^+-E & \Sigma_n\\
\Sigma_n^\ast & \Omega_{n+1}^--E
\end{array}
\right]
\left(
\begin{array}{c}
C_n^+\\
C_{n+1}^-
\end{array}
\right)
=0.
\end{equation}
Here $\Omega_n^\pm=\omega_0[(n+\phi)^2\mp 2(n+\phi)\bar{\alpha}\sin\gamma_1
+\bar{\alpha}^2+\bar{\beta}^2(\sin\gamma_2-\cos\gamma_2)^2]\pm\omega_B\cos\gamma_3$,
 $\Sigma_n=\hbar\omega_n(\bar{\alpha}\cos\gamma_1-i\bar{\beta}(\sin\gamma_2-\cos\gamma_2)
+\hbar\omega_B\sin\gamma_3$, and  $\omega_n=\omega_0(n+\phi+1/2)$.
In this representation the Hamiltonian is expressed as a matrix composed of a series
of $2\times2$ blocks and is explicitly Hermitian.
The secular equation is solved exactly.
The
eigenvalues (for $\sigma=\pm$) are
\begin{eqnarray}
\nonumber
E_{n\sigma}&=&\hbar\omega_0(n+\phi)^2
+\hbar\omega_0[\bar{\alpha}^2+\bar{\alpha}\sin\gamma_1+\bar{\beta}^2(\sin\gamma_2-\cos\gamma_2)^2]\\ \nonumber
&+&\hbar\omega_n
+\sigma\hbar(-\omega_n-2\omega_n\bar{\alpha}\sin\gamma_1+\omega_B\cos\gamma_3)
\cos\delta_{n\sigma}\\*
&+&\sigma\hbar[(2\omega_n\bar{\alpha}\cos\gamma_1+\omega_B\sin\gamma_3)^2
+4\omega_n^2\bar{\beta}^2(\sin\gamma_2-\cos\gamma_2)^2]^{1/2}
\sin\delta_{n\sigma}
\label{eng}
\end{eqnarray}
and the corresponding eigenvectors
\begin{equation}
\Psi_{n\sigma}=\frac{e^{i n\theta}}{\sqrt{2\pi}}\chi_{n\sigma}(\theta),\,\,\,
\text{ with spinors }\,\,\,
\chi_{n\sigma}(\theta)=
\left[
\begin{array}{c}
\cos(\delta_{n\sigma}/2)\\
\sin(\delta_{n\sigma}/2)e^{i\varphi_n+i\theta}
\end{array}
\right].
\label{wavf}
\end{equation}
Here $\delta_{n-}=\delta_{n+}-\pi$,
\begin{equation}
\cot(\delta_{n+})=\frac{-\omega_n(1+2\bar{\alpha}\sin\gamma_1)+\omega_B\cos\gamma_3}{
[(2\omega_n\bar{\alpha}\cos\gamma_1+\omega_B\sin\gamma_3)^2
+4\omega_n^2\bar{\beta}^2(\sin\gamma_2-\cos\gamma_2)^2]^{1/2}}\  ,
\end{equation}
and
\begin{equation}
\tan\varphi_n=\frac{\omega_n\bar{\beta}(\sin\gamma_2-\cos\gamma_2)}{
\omega_n\bar{\alpha}\cos\gamma_1+\omega_B\sin\gamma_3/2}\  .
\end{equation}
With the help of the Pauli matrices $\bm{\sigma}$ in  polar coordinates, we
easily find that the
orientation of the spin of the
state $\Psi_{n+}$
is $(\delta_{n+},\varphi_n+\theta)$
and opposite to that of the state $\Psi_{n-}$.

\begin{figure}[tpb]
\vspace{-7cm}
\includegraphics*[width=120mm]{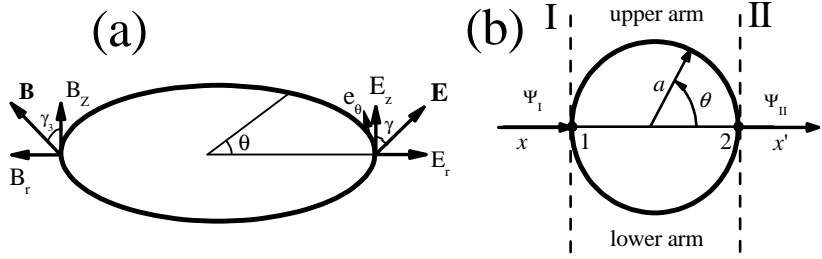}
\vspace{-6cm}
\caption{ (a) Geometry of an isolated
1D ring and the components of the electric and magnetic fields. (b) The ring of (a) connected to two
leads, regions I and II, and the corresponding wave functions.}
\label{fig1}
\end{figure}

If the ring is not isolated but coupled to outside leads, as shown in Fig. \ref{fig1},
the periodic boundary condition on $\theta$ is relaxed
and $n$ can be any real number. In this case, we find the above wave functions and eigenvalues
are still eigenvectors and eigenvalues of the Hamiltonian (\ref{ham}).
However, because $\delta_{n\sigma}$ and $\varphi_{n}$ depend
on $n$ in the presence of the Zeeman term,
the spin orientations of different eigenstates depend on $n$ and
the spinors $\chi_{n\sigma}(\theta)$
given by Eq. (\ref{wavf}) for electrons of the same energy are generally not orthogonal
to each other.
\cite{yi}

When the Zeeman term is negligible, $\delta_{n+}$ and $\varphi_{n}$
are independent of $n$ and $\sigma$ and will be replaced by $\delta$ and $\varphi$,
respectively.
In this case the energy  eigenvalues  (\ref{eng}) can be written as
\begin{equation}
E_{n\sigma}=\hbar \omega_0[n+\phi-\phi_{AC}^\sigma/(2\pi)]^2
\end{equation}
with the Aharonov-Casher phase $\phi_{AC}^\sigma$  given by
\begin{equation}
\phi_{AC}^\sigma=-\pi\{1+2\sigma[\bar{\alpha}^2+\bar{\alpha} \sin\gamma_1
+\bar{\beta}^2(\sin\gamma_2-\cos\gamma_2)^2+1/4]^{1/2}\}.
\end{equation}

Since the effects of the RSOI on the energy spectrum have been reported before, here we focus mostly on those on it
when the RSOI is replaced by the DSOI of a similar
strength. For clarity though we show its dependence on both $\alpha$ and $\beta$.
When $\bm{B}$,
$\bm{E}_R$, and $\bm{E}_D$ have only a $z$ component, the energy spectrum and
the angle $\delta$ versus the SOI strength $\alpha$ or $\beta$ do not change after this
replacement, though there is a rotation of the spin orientation in the $(x$-$y)$ plane by $\varphi$.
This is in line with the unitary equivalence between the  Rashba term and the linear Dresselhaus term \cite{sch}  that is  often exploited in the literature  \cite{ rash}. The radial components of $\bm{E}_R$ and $\bm{E}_D$, however, break  this equivalence and lead to different effects of the RSOI and DSOI on the energy spectrum as well as on $\delta$.

\begin{figure}[tpb]
\vspace{-5cm}
\includegraphics*[width=120mm]{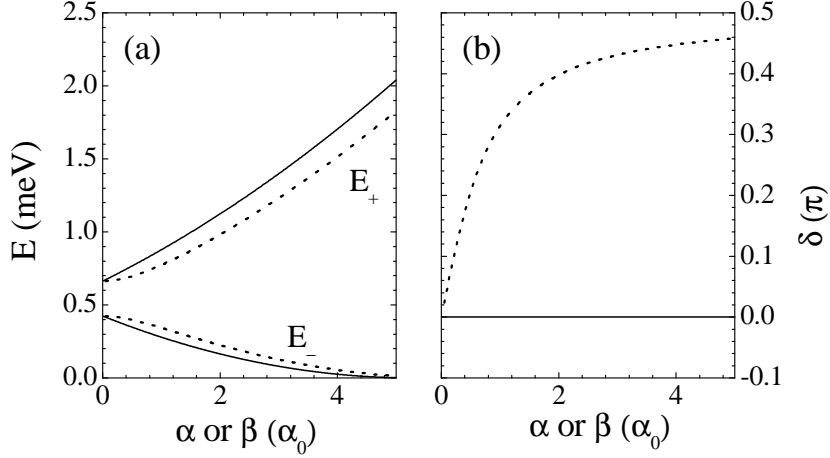}
\vspace{-4cm}
\caption{(a) Energy spectrum and (b) the angle between
the $z$ axis
and the
spin of the eigenstates of a 1D ring vs $\alpha$ or $\beta$
in the presence of only a radial SOI field
with
$\bm{E}_R=E_R\bm{e}_r$ (solid curves) and  $\bm{E}_D=E_D\bm{e}_r$ (dotted curves).
The solid curves in panel (a) also describe the energy spectrum
of the 1D ring as a function of  the combined SOI strength $\sqrt{\alpha^2+\beta^2}$
for $\bm{E}_R=E_R\bm{e}_z$ and
$\bm{E}_D=E_D\bm{e}_z$.
The magnetic field is perpendicular to the plane of the ring
and the Zeeman term is
neglected.
 The parameters used are  $n+\phi=4$, $g=0$,
$m^\ast=0.023$, $a=250$nm, $\gamma_1=\gamma_2=\pi/2$, and the strength unit
$\alpha_0=10^{-11}$ eVm.}
\label{fig2}
\end{figure}

In Fig. \ref{fig2} we show the energy spectrum and the absolute value of the angle $\delta$ for the
RSOI (solid curves) and
the DSOI (dotted curves). $\varphi$ equals  0 or $\pi$ in the former case and $\pi/2$ or $3\pi/2$ in the later.

\subsection{Transfer-matrix formulation}

{\it Single ring.} Now we consider the quantum transport of electrons with energy $E$
through a ring connected to two leads I and II
as shown in Fig. 1 with the local coordinate
systems attached to the different regions of the device.
Assuming there is no SOI in the leads,  the electron wavevector
is readily obtained as $k=\pm\sqrt{2m^\ast E}$.
Solving Eq. (\ref{eng}) for $n$, we obtain the angular wavevector
$n^\sigma_\mu$ of the electron with $\mu$  the mode index and  $\sigma$
the spin branch.
In this case it is appropriate to apply a
spin-dependent version of  Griffith's boundary conditions
\cite{grif,xia} at the intersections as  specified below. This reduces
the electron transport through the ring
to an exactly
solvable, 1D
scattering problem.  The
conditions at each junction are: (i) the wave function must be
continuous, and (ii) the spin probability current density must be
conserved. Notice that we consider the case
where the magnetic field is weak and the Zeeman term is negligible.
When the Zeeman term is taken into account exactly,
the resulting spinors are not orthogonal; this renders
the transmission
through each junction
and the ring uncertain  and a
phenomenological parameter is required to solve the problem. \cite{yi}

The incident wavefunction $\Psi_I$ and the outgoing one $\Psi_{II}$ can be expanded
in terms of spinors $\chi_{n\sigma}$
\begin{eqnarray}
\Psi_I(x)& =\sum_\sigma\Psi^\sigma_I(x)=\sum_\sigma (A^\sigma e^{ikx}+ B^\sigma e^{-ikx})
\chi ^\sigma (\pi ),\hspace{0.85cm}x\in \left[ -\infty ,0\right] , \\
\Psi_{II}(x^\prime)&=\sum_\sigma\Psi^\sigma_{II}(x^\prime) =\sum_\sigma
(C^\sigma e^{ikx^\prime}+ G^\sigma e^{-ikx^\prime})
\chi ^\sigma (0 ),
\hspace{0.85cm}x^{\prime }\in \left[
0,\infty \right],
\end{eqnarray}
see Fig. 1 for the local coordinates $x$ and $x^{\prime }$.
In a similar way the wave
functions corresponding to the upper and lower arms of the ring can be
written as
\begin{eqnarray}
\Psi_u(\theta)& =\sum_\sigma\Psi^\sigma_u(\theta)
=\sum_{\sigma,\mu}D_\mu e^{in_\mu^\sigma\theta}
\chi^\sigma(\theta),\hspace{0.85cm}\theta \in \left[ 0,\pi \right],
\\
\Psi_l(\theta)& =\sum_\sigma\Psi^\sigma_l(\theta)=\sum_{\sigma,\mu}F_\mu e^{in_\mu^\sigma\theta}
\chi^\sigma(\theta),\hspace{0.85cm}\theta \in \left[\pi,2\pi \right].
\end{eqnarray}
In the absence of the Zeeman term,
energy conservation leads to
$n_\mu^\sigma=\mu ka 
-\phi+\phi^\sigma_{AC}/(2\pi)$ with $\mu=\pm 1$.

Using  Griffith's boundary conditions described above and the spin current operator
\begin{equation}
J^\sigma=\frac{\hbar}{m^\ast}\text{Re}
\{(\Psi^\sigma)^\dag
(-i 
\partial/\partial \theta +
\hat{\sigma})\Psi^\sigma \}
\end{equation}
with $\hat{\sigma}=a[\bar{\alpha} \cos\gamma_1\sigma_r
-\bar{\alpha}\sin\gamma_1\sigma_z
+\bar{\beta}(\sin\gamma_2-\cos\gamma_2)\sigma_\theta]/\hbar+\phi$,
we have
\begin{eqnarray}
\Psi^\sigma_I(0)=\Psi^\sigma_u(\pi)=\Psi^\sigma_l(\pi),\\
\Psi^\sigma_{II}(0)=\Psi^\sigma_u(0)=\Psi^\sigma_l(2\pi),\\
-ia\frac{\partial}{\partial x}\Psi^\sigma_I(x) |_{x=0}+
(-i\frac{\partial}{\partial \theta}+\hat{\sigma})\Psi^\sigma_u(\theta)|_{\theta=\pi}
-(-i\frac{\partial}{\partial \theta}+\hat{\sigma})\Psi^\sigma_l(\theta)|_{\theta=\pi}=0,\\
-ia\frac{\partial}{\partial x}\Psi^\sigma_{II}(x^\prime) |_{x^\prime=0}+
(-i\frac{\partial}{\partial \theta}+\hat{\sigma})\Psi^\sigma_u(\theta)|_{\theta=0}
-(-i\frac{\partial}{\partial \theta}+\hat{\sigma})\Psi^\sigma_l(\theta)|_{\theta=2\pi}=0.
\end{eqnarray}
Then
the coefficients of the wave functions in regions I and II are related by
\begin{equation}
\left(
\begin{array}{c}
A^\sigma \\
B^\sigma
\end{array}
\right)=\frac{1}{2ka}
\left[
\begin{array}{cccc}
k_{1-}e^{in_1^\sigma\pi} &
\ \ k_{2-} 
e^{i\pi n_2^\sigma} &
\ \ n_1^\sigma e^{-i\pi n_1^\sigma } &
n_2^\sigma e^{-in_2^\sigma\pi} \\
k_{1+} 
e^{in_1^\sigma\pi} &
\ \ k_{2+} 
e^{i\pi n_2^\sigma} &
\ \ -n_1^\sigma e^{-i\pi n_1^\sigma } &
\ \ -n_2^\sigma e^{-in_2^\sigma\pi}
\end{array}
\right]
\left(
\begin{array}{c}
D_{1}^\sigma \\
D_{2}^\sigma \\
F_{1}^\sigma \\
F_{2}^\sigma
\end{array}
\right)
\label{mp1}
\end{equation}
and
\begin{equation}
\hat{P}^\sigma
\left(
\begin{array}{c}
D_{1}^\sigma \\
D_{2}^\sigma \\
F_{1}^\sigma \\
F_{2}^\sigma
\end{array}
\right)
=\left[
\begin{array}{cc}
0&0 \\
1&1 \\
1&1 \\
-ak&ak
\end{array}
\right]
\left(
\begin{array}{c}
C^\sigma \\
G^\sigma
\end{array}
\right)
\label{mp2}
\end{equation}
with
\begin{equation}
\hat{P}^\sigma=\left[
\begin{array}{cccc}
e^{in_1^\sigma\pi} & e^{i\pi n_2^\sigma} & -e^{-i\pi n_1^\sigma } &
-e^{-in_2^\sigma\pi} \\
1 & 1 & 0 & 0 \\
0 & 0 & e^{i2\pi n_1^\sigma} & e^{i2\pi n_2^\sigma}\\
n_1^\sigma & n_2^\sigma & -n_1^\sigma e^{i2\pi n_1^\sigma} &
-n_2^\sigma e^{-i2\pi n_2^\sigma}
\end{array}
\right],
\label{mp3}
\end{equation}
where $k_{s\pm} =(ak\pm n_s^\sigma), s=1,2$.
Now we can write the transfer matrix $M^\sigma$ through a single ring in the representation
of the eigenspinors of the ring at $\theta=\pi$ and $\theta=0$ for the incident and outgoing
wave function, respectively.

If the Zeeman term is neglected, the transmission amplitude through a single ring,
with both terms of the SOI present, takes the same form
as that in the absence of the DSOI \cite{moln}
\begin{equation}
t_\sigma=C^\sigma/A^\sigma=\frac{8i\cos (-\phi/2+\phi _{AC}^\sigma/2)\sin (ka\pi )}{1-5\cos (2ka\pi
)+4\cos(-\phi+\phi _{AC}^\sigma)+4i\sin (2ka\pi )}
\label{tlast}
\end{equation}
but with the  phase
$\phi _{AC}^\sigma$ modified, cf. Eq. (15), and accounting for $\beta$ and the angles $\gamma_{1}, \gamma_{2}$. Here $\sigma=\pm$ corresponds to the $\pm$ spinors at $\theta=\pi$ and $\theta=0$
of the ring for incident and output electrons respectively.
At zero temperature, the conductance of the ring is given by
\begin{equation}
G=\frac{e^{2}}{h}\sum_\sigma | t_\sigma | ^{2}
=\frac{e^{2}}{h}\sum_\sigma g_0(k,\Delta^\sigma_{AC})[1-\cos (\Delta^\sigma _{AC})].
\label{cont}
\end{equation}
The dimensionless coefficent $g_{0}$ is given by
\begin{equation}
g_{0}(k,\Delta^\sigma_{AC})=\frac{32\sin ^{2}(ka\pi )}{[1-5\cos (2ka\pi )-4\cos
(\Delta^\sigma_{AC})]^{2}+16\sin ^{2}(2ka\pi )}
\label{Con1}
\end{equation}
and $\Delta^\sigma_{AC}=-\phi+\sigma(\phi_{AC}^+-\phi_{AC}^-)/2$.

{\it A series of  rings}.
For single a ring we used different spinors
$\chi^\sigma(\pi)$ and $\chi^\sigma(0)$ for the incident and outgoing electrons.
This introduces inconveniences to the description of spins because, e.g.,  the "$+$ spinor" may
represent different
spin orientations for the incident and outgoing electrons. Furthermore,
if there are more than one rings in the system,
in order to take advantage of the single-ring results, a unitary transformation is necessary
between the outgoing (from one ring)
and the incident (to the next ring) spinor representations.
Accordingly,
in the following we will work in the representation of the eigenspinors
of $\sigma_z$, $(1,0)^T$ for spin-up states (branch $+$) and $(0,1)^T$
for spin-down states (branch $-$). Here the superscript $T$ denotes the
transpose of a matrix. We use $U_L$ and $U_R$ to denote the unitary transformation matrices between
the $\sigma_z$ representation and that for the incident ($\chi^\sigma(\pi)$) and outgoing ($\chi^\sigma(\pi)$)
wave functions, respectively.
We express the incident
wave functions as
\begin{eqnarray}
&\Psi_I(x)=
\left(
\begin{array}{c}
\mathcal{A}^+\\
\mathcal{A}^-
\end{array}
\right)
e^{ikx}+
\left(
\begin{array}{c}
\mathcal{B}^+\\
\mathcal{B}^-
\end{array}
\right)
e^{-ikx};
\end{eqnarray}
the outgoing wave function $\Psi_{II}(x)$ is given by the same expression
with $\mathcal{A}$ and $\mathcal{B}$ replaced by $\mathcal{C}$ and $\mathcal{G}$,
respectively.  The coefficients $\mathcal{C}$ and $\mathcal{G}$
are related to $C$ and
$G$ in the manner
\begin{equation}
\left(
\begin{array}{c}
D^+\\
D^-
\end{array}
\right)
=U_R
\left(
\begin{array}{c}
\mathcal{D}^+\\
\mathcal{D}^-
\end{array}
\right),\,\,\,
U_R=
\left[
\begin{array}{cc}
\cos(\delta/2)& e^{i\varphi} \sin(\delta/2)\\
\sin(\delta/2)& -e^{i\varphi} \cos(\delta/2)
\end{array}
\right],
\end{equation}
where $D=C, G$. Similarly, the coefficients
$\mathcal{A}$ and $\mathcal{B}$ are related to $A$ and
$B$ in the manner
\begin{equation}
\left(
\begin{array}{c}
\mathcal{E}^+\\
\mathcal{E}^-
\end{array}
\right)
=U_L
\left(
\begin{array}{c}
E^+\\
E^-
\end{array}
\right),\,\,\,
U_L=
\left[
\begin{array}{cc}
\cos(\delta/2)& \sin(\delta/2)\\
-e^{-i\varphi} \sin(\delta/2)& e^{-i\varphi} \cos(\delta/2)
\end{array}
\right].
\end{equation}
with $E=A,B$. Then the transfer matrix of the $i$th ring in the $\sigma_z$ representation  becomes
\begin{equation}
\left(
\begin{array}{c}
\mathcal{A}^+\\
\mathcal{A}^-\\
\mathcal{B}^+\\
\mathcal{B}^-
\end{array}
\right)=\mathcal{M}_i
\left(
\begin{array}{c}
\mathcal{C}^+\\
\mathcal{C}^-\\
\mathcal{G}^+\\
\mathcal{G}^-
\end{array}
\right)
\label{tranm}
\end{equation}
with
\begin{equation}
\mathcal{M}_i=
\left[
\begin{array}{cc}
U_L&0 \\
0&U_L
\end{array}
\right]
\left[
\begin{array}{cccc}
M_{11}^+&0&M_{12}^+&0 \\
0&M_{11}^-&0&M_{12}^- \\
M_{21}^+&0&M_{22}^+&0 \\
0&M_{21}^-&0&M_{22}^-
\end{array}
\right]
\left[
\begin{array}{cc}
U_R&0 \\
0&U_R
\end{array}
\right]
\end{equation}
and $M_{ij}^\sigma$ the elements of the transfer matrix $M^\sigma$
determined by Eqs. (\ref{mp1})-(\ref{mp3}).

For a system of $n$ rings,
the matrix $\mathcal{M}_i$
in Eq. (\ref{tranm}) should be replaced by the  total transfer matrix of the system
\begin{equation}
\mathcal{M}_T=\prod_i \mathcal{M}_i
\end{equation}

If  no electrons enter  the system from the right lead, i.e., if
$\mathcal{G}^+=\mathcal{G}^-=0$, the spin transmission and reflection rates
can be calculated by relating
the coefficients of the transmitted
$(\mathcal{C}^+,\mathcal{C}^-)^T$ and
  reflected $(\mathcal{B}^+,\mathcal{B}^-)^T$ wave function  to those of
the incident wave function $(\mathcal{A}^+,\mathcal{A}^-)^T$ in Eq.(\ref{tranm}).
The  total transmission is obtained as
\begin{equation}
T=\sum_\sigma T^\sigma=\sum_\sigma |\mathcal{C}^\sigma|^2/(|\mathcal{A}^+|^2+|\mathcal{A}^-|^2)
\label{tran}
\end{equation}
with
\begin{equation}
\left(
\begin{array}{c}
\mathcal{C}^+\\
\mathcal{C}^-
\end{array}
\right)
=
\left[
\begin{array}{cc}
\mathcal{M}_{11}&\mathcal{M}_{12} \\
\mathcal{M}_{21}&\mathcal{M}_{22}
\end{array}
\right]^{-1}
\left(
\begin{array}{c}
\mathcal{A}^+\\
\mathcal{A}^-
\end{array}
\right)
\end{equation}
and $\mathcal{M}_{ij}$ the elements of the total matrix $\mathcal{M}_T$.
In the following numerical calculation, we
take the incident spinor as
$\mathcal{A}^+=\mathcal{A}^-=\sqrt{2}/2$ which
is oriented along the $x$ direction.
The zero-temperature conductance is then evaluated from
the transmission of electrons at the Fermi energy as $G=2e^2 T/h$.
Note the partial transmissions $T^\sigma$ in Eq. (\ref{tran})
and the partial transmission amplitude $t_\sigma$ in
Eq. (\ref{tlast}) are defined in
different spinor representations but
they give the same
conductance
through a single ring.

\section{Results and Discussion}
\subsection{Single ring}

\begin{figure}[tpb]
\vspace{-2cm}
\includegraphics*[width=70mm]{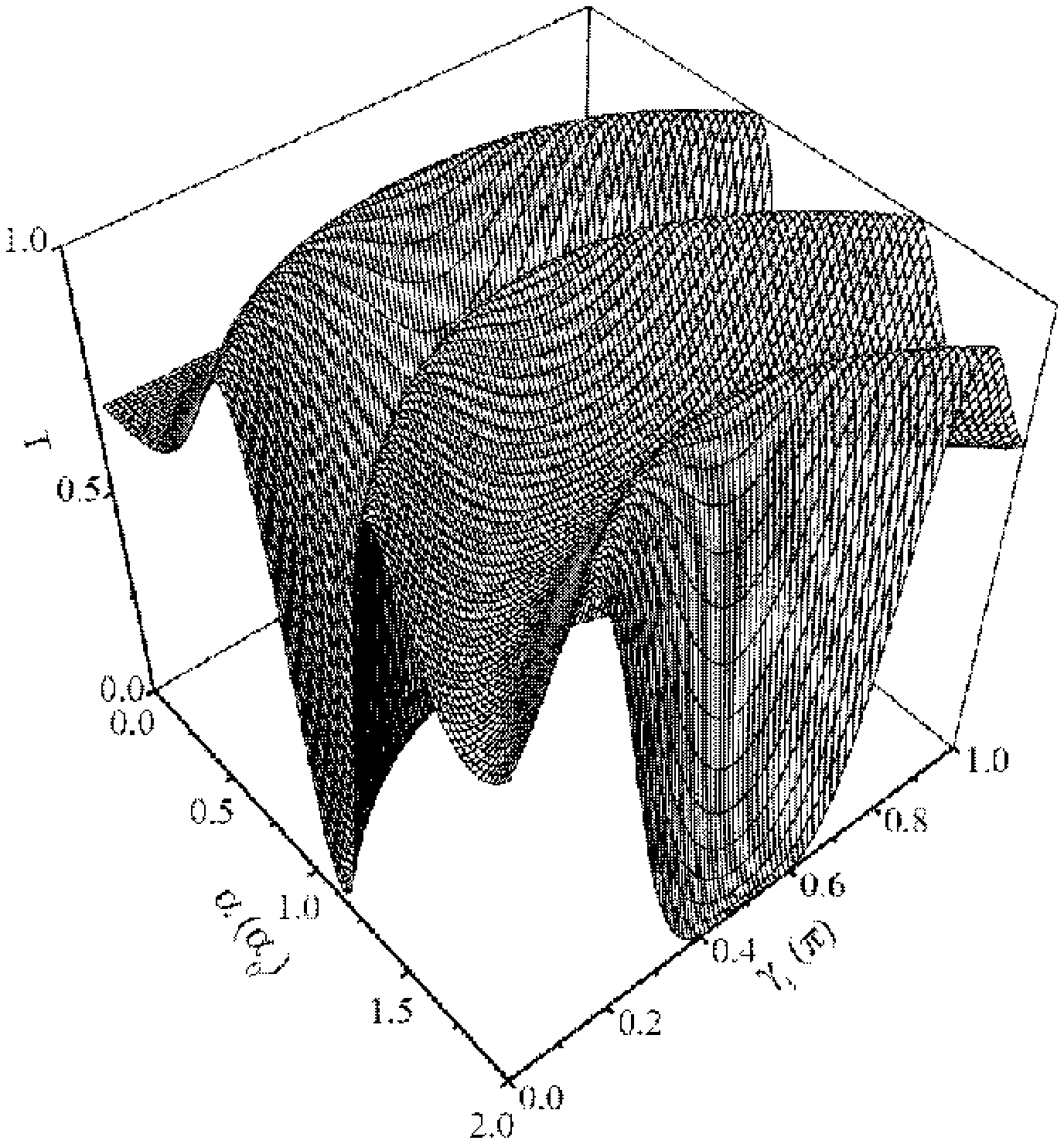}
\includegraphics*[width=70mm]{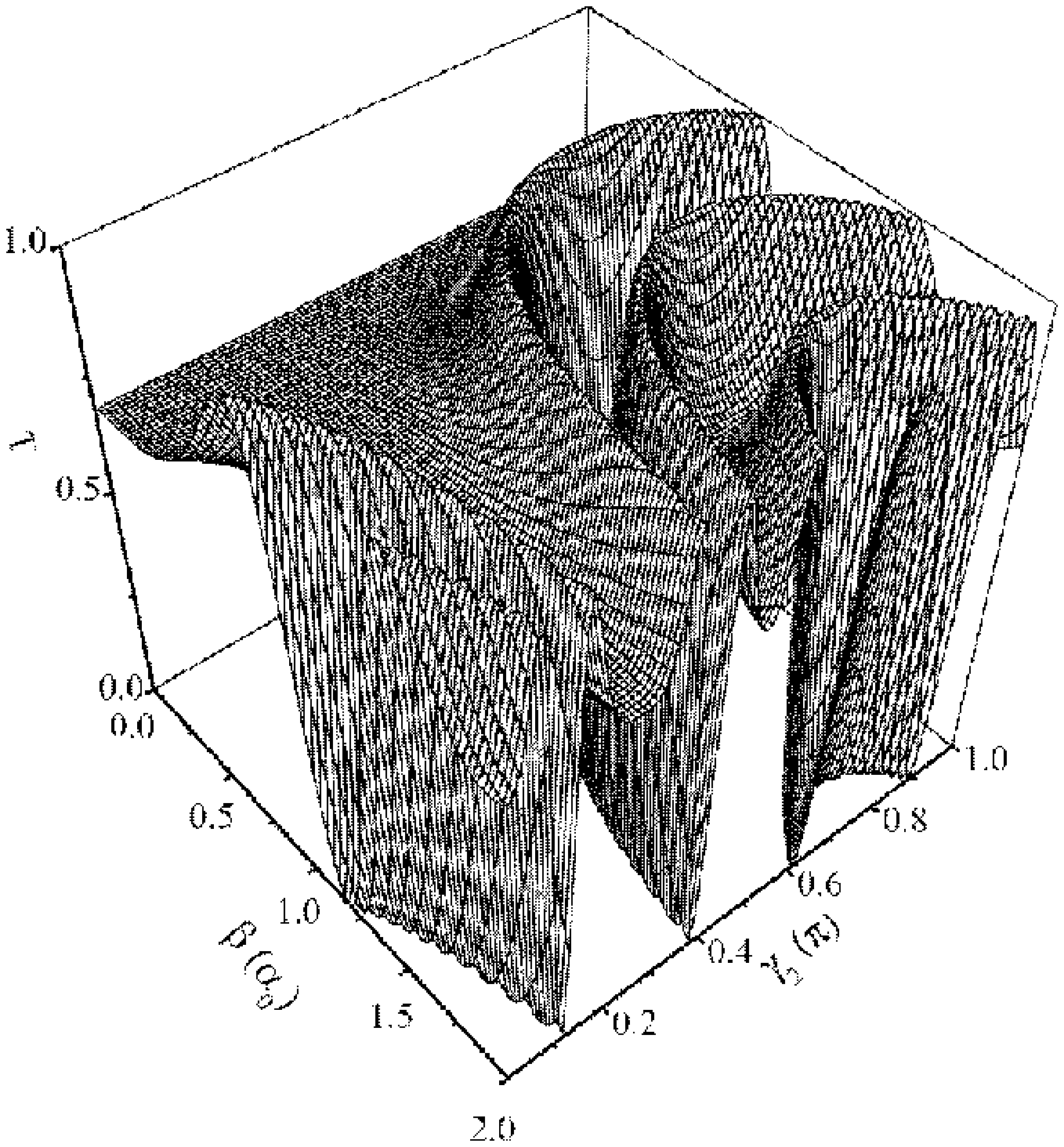}
\vspace{-0cm}
\caption{Total transmission through a ring vs $\alpha$ and $\gamma_1$ in (a), for $\beta=0$,
and vs $\beta$ and $\gamma_2$ in (b) for  $\alpha=0$.
The other parameters are $B=0$, $a=25nm$, $m^\ast=0.023$, and $E=11.3$meV.}
\label{fig3}
\end{figure}
In Fig. \ref{fig3}, we show the transmission as function of the SOI strength and of the SOI electric field orientation, for $\beta=0$
in (a) and for  $\alpha=0$ in (b),
in the absence of a magnetic field.
In (a) the transmission is symmetric along the angle $\gamma_1$,
with respect to the line $\gamma_1=\pi/2$.
In (b) on the contrary, the transmission through the ring depends on
the sign of the $z$ component of $\bm{E}_D$.
As illustrated in the Hamiltonian  (\ref{ham}), the RSOI affects the system via $\sigma_r$
and $\sigma_z$ but the DSOI via only $\sigma_\theta$.
As a result, the energy spectrum is symmetric
along $\gamma_1$ with respect to $\gamma_1=\pi/2$ in the absence of the Zeeman term
but asymmetric along $\gamma_2$. This explains the different behavior of the transmission
as a function of the RSOI parameters $\alpha, \gamma_1$ or of the DSOI  parameters $\beta, \gamma_2$ shown in Fig. \ref{fig3}.
\begin{figure}[tpb]
\vspace{-0cm}
\includegraphics*[width=70mm]{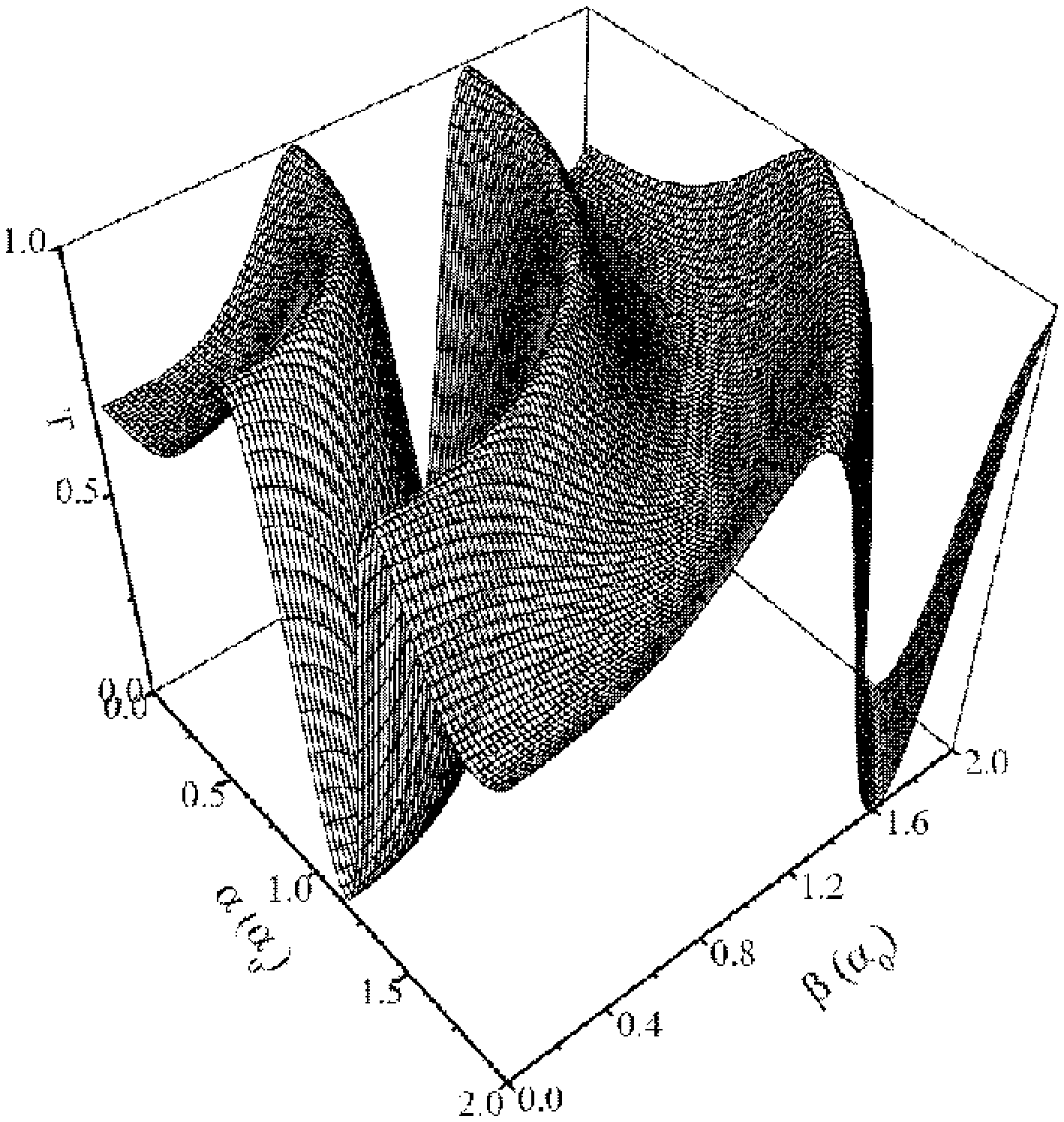}
\includegraphics*[width=70mm]{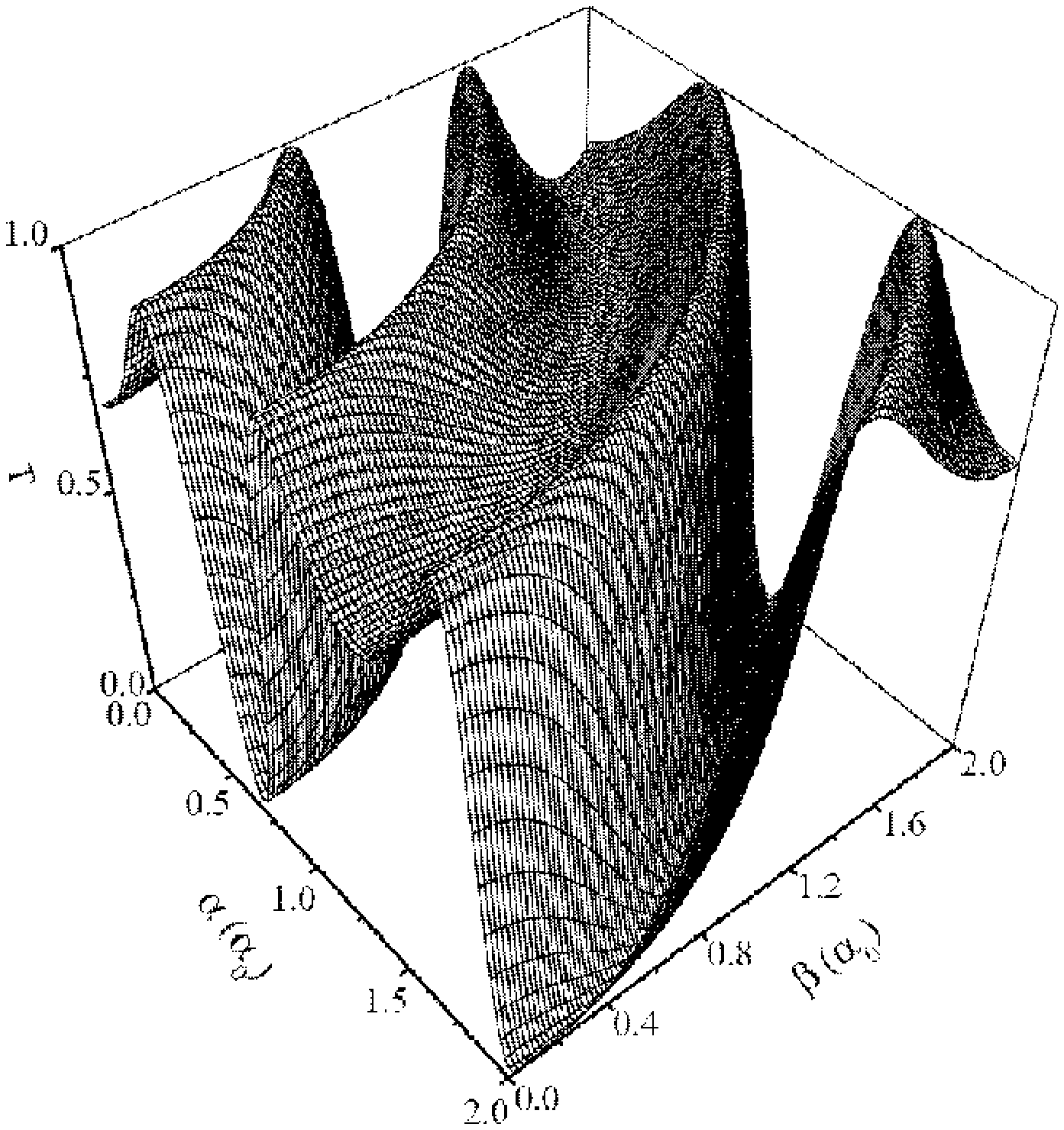}
\vspace{-0cm}
\caption{ Total transmission through a ring vs $\alpha$ and $\beta$ for $\gamma_1=\gamma_2=0$ in (a) and
$\gamma_1=\gamma_2=\pi/2$ in (b). The other parameters are
$B=0$, $a=25nm$, $m^\ast=0.023$, and $E=11.3$meV.}
\label{fig4}
\end{figure}

In Fig. \ref{fig4}, the transmission in the ($\alpha-\beta$) space is shown
for  $\gamma_1=\gamma_2=0$ in panel (a) and  $\gamma_1=\gamma_2=\pi/2$ in panel (b).
Similar to the energy spectrum, the transmission is a function of $(\alpha^2+\beta^2)^{1/2}$
and shows a symmetry along the $\alpha$ and $\beta$ axes if both $\bm{E}_R$ and $\bm{E}_D$
are along the $z$ direction.
 This is in line with the unitary equivalence of the RSOI and DSOI in a 2DEG for
$\gamma_1=\gamma_2=0$, as discussed in Ref. \cite{sch}, and can be deduced from Eq. (7) . However, this is not the case for $\gamma_1\neq\gamma_2\neq 0$. For instance, if both $\bm{E}_R$ and $\bm{E}_D$
are along the radial direction, the transmission
along the $\alpha$ axis is different than that
along the $\beta$ axis. As a result,
the curves in the ($\alpha-\beta$) plane of equal transmission are circles in Fig. \ref{fig4}(a)
and ellipses in Fig. \ref{fig4}(b).

\begin{figure}[tpb]
\vspace{-2cm}
\includegraphics*[width=70mm]{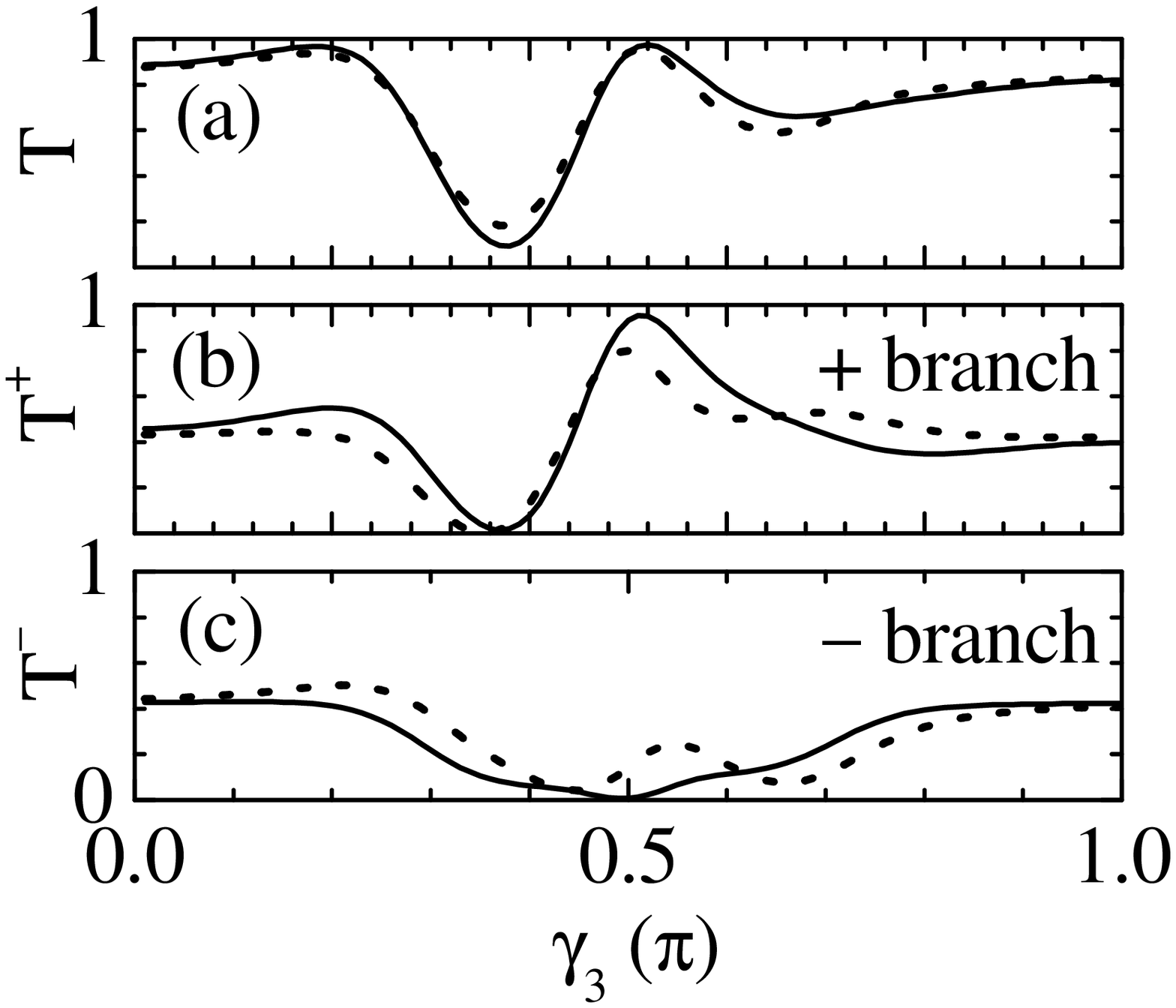}
\vspace{-2cm}
\caption{Total transmission (a), branch $+$ transmission $T^{+}$ (b), and branch $-$ transmission $T^{-}$ (c)
through a ring vs $\gamma_3$ under a cone-shaped magnetic field of magnitude $0.01$T.
The solid curves are for  $\alpha=\sqrt{2}\alpha_0$ and the dotted ones for
$\alpha=\beta=\alpha_0$.
The other parameters are $a=25nm$, $m^\ast=0.023$, $E=11.3$meV, and $\gamma_1=\gamma_2=0$.}
\label{fig5}
\end{figure}

In Fig. 5 we show the transmission vs the azimuthal angle $\gamma_3$ of
a weak magnetic field $B=0.01$T. The transmission oscillates upon varying this angle.
When $\gamma_3=0$
the total transmissions for $\alpha=\beta=\alpha_0$ and
for $\alpha=\sqrt{2}\alpha_0$ and $\beta=0$
are identical but there is a small difference between the partial transmissions
for the two sets of parameters. In a tilted magnetic field ($\gamma_3\neq 0$),
the rotational symmetry of the transmission in the
($\alpha$-$\beta$) space, shown in Fig. \ref{fig4}(a), is broken and there is a
a  difference between the solid and dotted curves in Fig. \ref{fig5}.

\subsection{Multiple rings}

\begin{figure}[tpb]
\vspace{-2cm}
\includegraphics*[width=70mm]{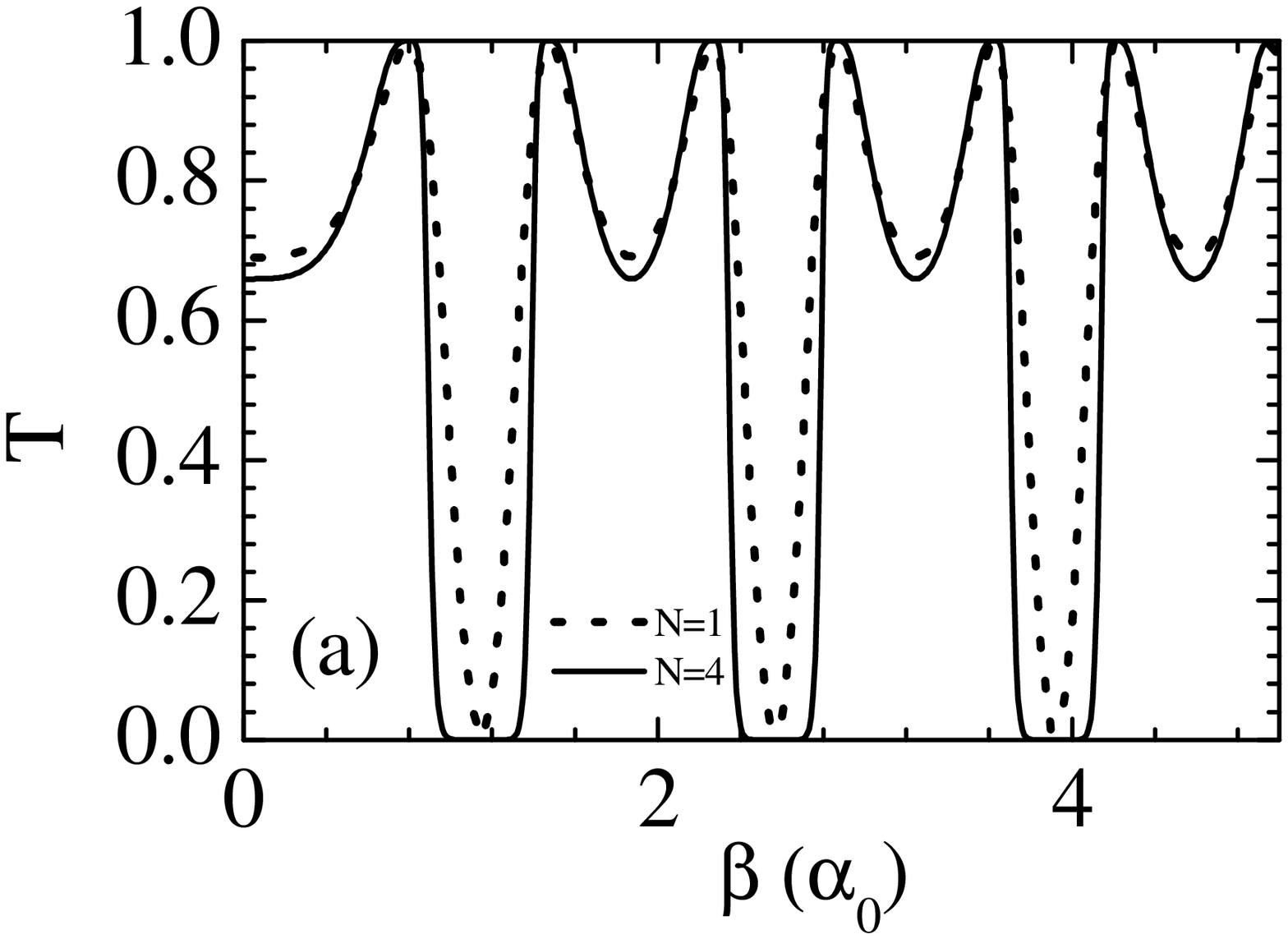}
\includegraphics*[width=70mm]{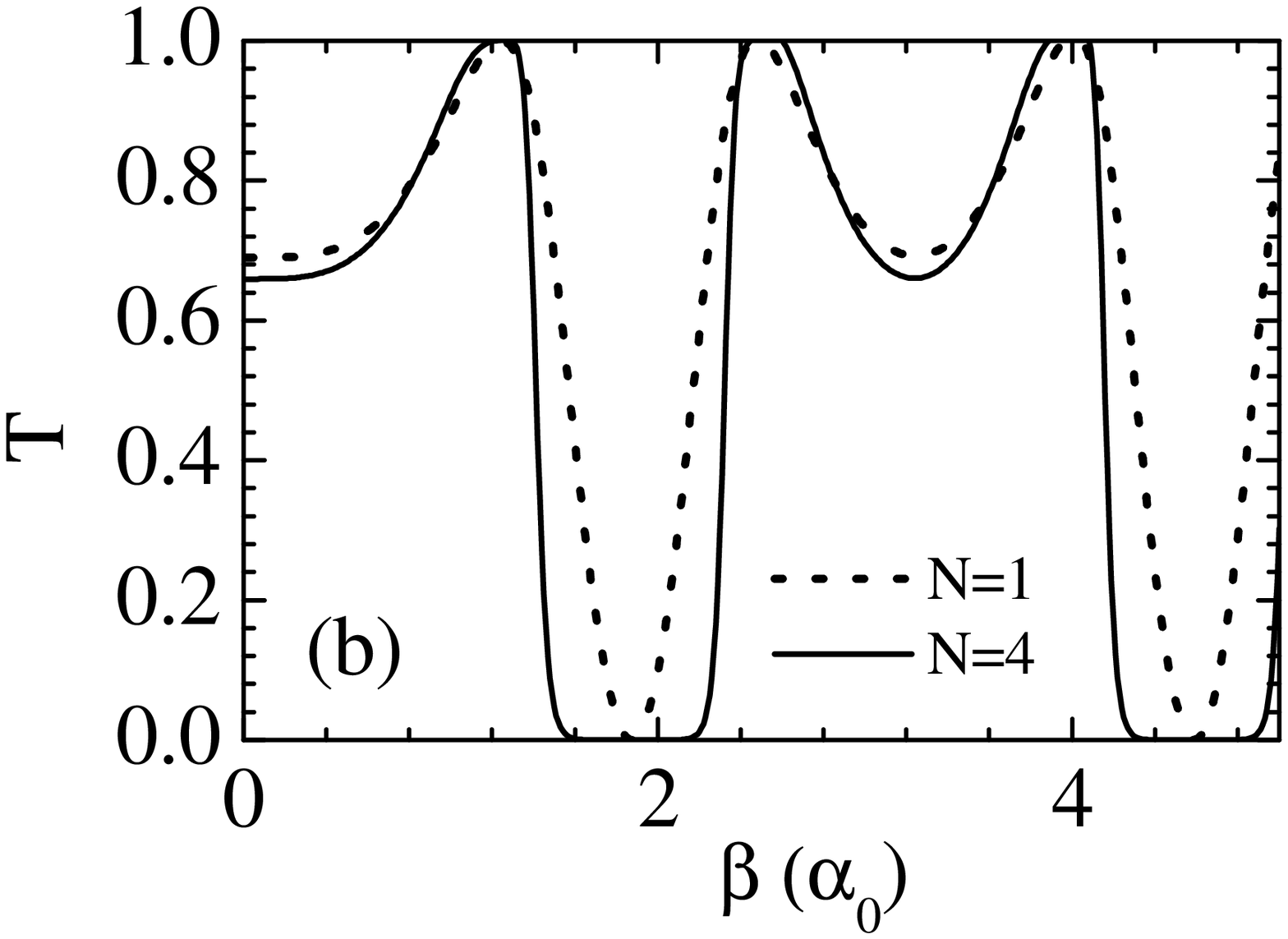}
\vspace{-2cm}
\caption{Total transmission through ring structures vs $\beta$ when the DSOI exists (a) in both
arms of the rings or (b) only in their lower
arms. The dotted curves show the result for one ring and the solid
ones  the result for  four rings.
The other parameters are $a=25nm$, $m^\ast=0.023$, $E=11.3$meV, $\alpha=0$, $B=0$,
 and $\gamma_1=0$.}
\label{fig6}
\end{figure}

For a series of identical rings
the transmission gaps become wider and acquire a square-wave character. This
is evident  in Fig. \ref{fig6}, where results
for one and four rings are shown for the case that the SOI exists
everywhere in the rings (a) or only in their lower arms (b).
When the SOI exists in both arms, it introduces
a phase difference between electrons propagating in the upper and  lower arms.
This is greatly reduced in a system where the SOI exists  only in one of the arms.
As a result, the  oscillation frequency of the transmission,
when the SOI strength is varied, is reduced. In Fig. \ref{fig6}
we show results only for the DSOI but similar results are obtained for the RSOI.

\begin{figure}[tpb]
\vspace{-2cm}
\includegraphics*[width=100mm]{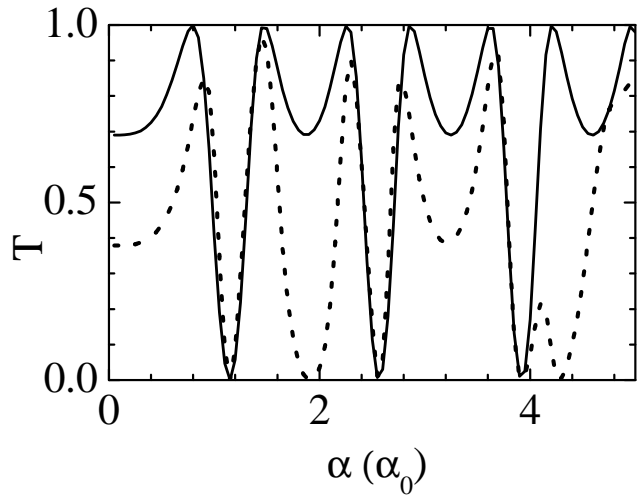}
\vspace{-3cm}
\caption{Transmission through two rings as a function of the
strength $\alpha$ for two sets of radii: $a_1=a_2=0.25
\mu$m (solid  curve) and $a_1=0.25\ \mu$m, $a_2=0.15\  \mu$m (dashed  curve).}
\label{fig7}
\end{figure}

For a series of rings, we can also change the ring radius $a$ from one ring to
another. This leads to a significant   change in the transmission pattern and
results from the modification of the energy levels, when $a$ changes,
cf. Eqs. (\ref{ham}) and (\ref{eng}).
In Fig. \ref{fig7} we show results for two rings of the same radius
(solid curve) and of different radii (dotted curve). For a series
of many rings, the gaps acquire a more pronounced square-type character,
see Ref. \cite{mol1} for more results when
$\beta$, $\gamma_1$, and $\gamma_3$
are zero.

If only the RSOI term is present and $\gamma_1$ vanishes, a previous study \cite{mol1} showed that for $ka=(2m+1)/2$, with $m$ integer, and
{\it even} number of identical rings,
the zero-magnetic-field transmission is a discontinuous  function of $\alpha$:
it takes the highest constant value and vanishes
at some special points $\alpha_{m}$  given by
$\alpha_{m}=(\hbar^2/2m^\ast a) \sqrt{4(m+1)^2-1}$.
On the other hand, the transmission through an {\it odd} number of rings   is identical to that for  one ring. The analytical explanation relies on the properties of the corresponding transfer-matrix that connects the expansion coefficients in leads I and II
 \cite{mol1}.  We verify numerically that the same holds in our much more complex situation.

\begin{figure}[tpb]
\vspace{-2cm}
\includegraphics*[width=100mm, height=190mm]{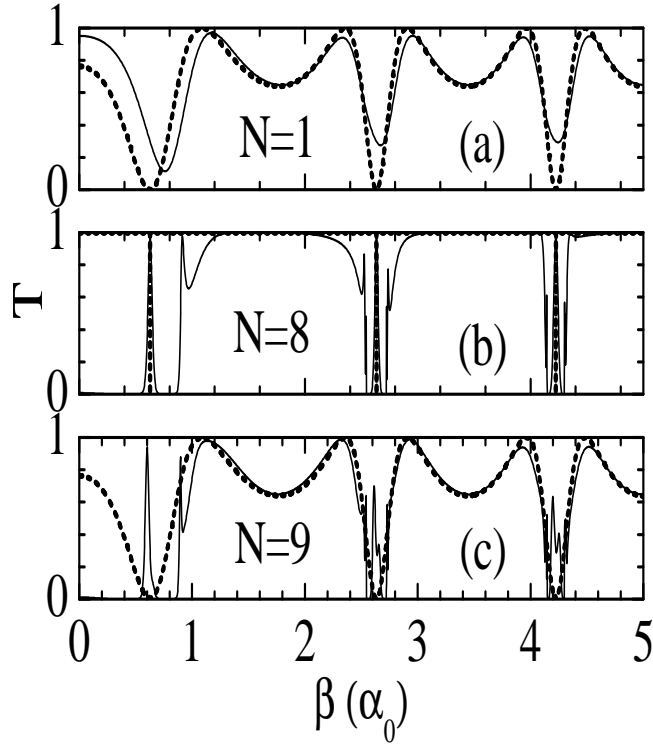}
\vspace{-4cm}
\caption{Transmissions for half-integer $ka=20.5$ through $N=1, 8, 9$ rings,
 as functions of the
strength $\beta$ without magnetic field (dotted curves) and with magnetic field of
magnitude $B=0.001$T (solid curves).
The parameters are $a=0.25\mu$m, $E=11.138$meV, $\alpha=\alpha_0$, $\gamma_1=0$, and $\gamma_2=\pi/32$.}
\label{fig8}
\end{figure}

For a ring of radius $a=0.25\mu$m and with only RSOI,
the transmission of an even number of rings vanishes
at $\alpha=1.148\alpha_0$, $2.566\alpha_0$,
and $3.92\alpha_0$ corresponding to $m$=1, 2, and 3.
If only the DSOI is present, the  transmission exhibits the
same profile  if $\alpha$ is replaced by $\beta$.
In a system with both DSOI and RSOI present, similar results hold.
If we tilt the SOI fields by increasing $\gamma_1$ and $\gamma_2$ from zero,
the separation between these points increases.
We show that in  Fig. \ref{fig8} where we plot the transmissions (thick dotted
curves) as  functions of  $\beta$ for $N=1, 8, 9$ rings.
The RSOI term is for $\alpha=\alpha_0$ and $\gamma_1=0$
and the tilt angle of the DSOI field is $\gamma_2=\pi/32$.
For  $N=8$ the transmission vanishes  at
$\beta=0.628\alpha_0$, $2.634\alpha_0$, and $4.225\alpha_0$.
The transmission through $N=8$ rings vanishes
at the same points as that for one ring and otherwise is equal to one.
The transmission through $N=9$ rings is identical to that
for  one ring. The dotted curves in Fig. 8 make these two statements clear
if we compare the curves for $N= 1$ and $N= 8$ and separately for $N=1$ and $N=9$.
This holds only if the magnetic field is zero.
If we apply a finite magnetic field to the system, the above result breaks down and
the transmission, as shown by the thin solid curves in Fig. \ref{fig8},
oscillates near the points at which it vanishes when the  magnetic field is zero.

\subsection{Influence of the Zeeman term}
\begin{figure}[tpb]
\vspace{-2cm}
\includegraphics*[width=100mm]{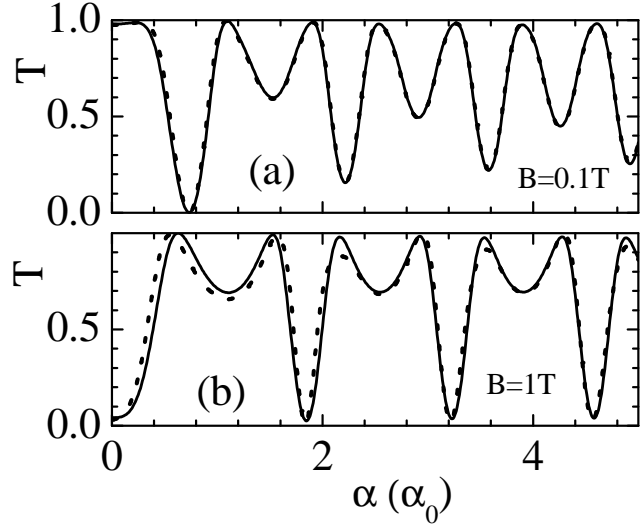}
\vspace{-3cm}
\caption{Transmission  through a ring vs the RSOI strength $\alpha$ when (a)
a weak magnetic field $B=0.1$ T and (b) a medium magnetic field $B=1$ T is applied. The solid curves are the results
without the Zeeman term   and the dotted ones with it  
($g=10$). The other parameters are $a=0.25 \mu$m, $m^\ast=0.023$, $E=11.3$meV, $\beta=0$,
and $\gamma_1=0$. }
\label{fig9}
\end{figure}

So far we neglected the effect of the Zeeman term on the transmission.
The reason is that in the presence of this term
the spin orientations of different eigenstates depend on the orbital quantum number $n$ and
for electrons of the same energy
the spinors
are not orthogonal to each other \cite{yi}.
This renders the transmission unwieldy or very difficult
to solve unless one resorts to the treatment
of Ref.  \cite{yi} at the expense of introducing a phenomenological parameter.
However, the question  arises to what extent its inclusion would modify the previous results.
An exact numerical treatment is beyond the scope of this work but it is probably unnecessary
for weak magnetic fields $B\leq 1$ T for which its influence  can by assessed
by treating it as a perturbation and neglecting the correction to the eigenfunctions (10).
The resulting eigenvalues read

\begin{equation}
E_{n\sigma}=\hbar\omega_0(n+\phi)^2+\hbar\omega_0(\bar{\alpha}^2+\bar{\beta}^2)
+\hbar\omega_n+\sigma\hbar(\omega_B-\omega_n)
+2\sigma\omega_n(\bar{\alpha}^2+\bar{\beta}^2)/(\bar{\alpha}^2+\bar{\beta}^2+1/4)^{1/2}.
\end{equation}
For a typical SOI strength  $\sqrt{\alpha^2+\beta^2}=\alpha_0$,
the ratio between the spin splitting due to the Zeeman term and that due the
SOI is $\omega_B\sqrt{\alpha_0^2+1/4}/2\alpha_0^2\omega_n$.
Notice that $\omega_n$ is of the same order as the electron energy $E$.
For the parameters used in Fig. \ref{fig9} and $B=1$ T  this ratio is about $2\%$.
In Fig. 8  the transmission
without ($g=0$) and with ($g=10$) the Zeeman term is shown  for perpendicular magnetic fields
$B=0.1$ T in (a) and $B=1$ T in (b). Upon increasing the weak magnetic field
some of the transmission gaps vary
due to the change in the A-B phase. As shown though,
the Zeeman term has an
overall negligible effect. Notice though that for  $\alpha$  close to zero  the perturbation
treatment is not valid despite the agreement between the results without  and with the Zeeman term.

\section{ Concluding remarks}

We studied the electron energy spectrum and transmission  through mesoscopic rings, in a tilted magnetic field, in the presence of the
Rashba (RSOI) and Dresselhaus (DSOI) terms of the spin-orbit interaction
due to the confinement along the perpendicular
and   radial direction. We solved exactly the one-electron Schr\"{o}dinger equation and obtained the spectrum and the AC phase including the Zeeman term. This was followed by the formulation of the transmission problem, using a  spin-dependent version of Griffith's boundary conditions, with  the Zeeman term treated by perturbation theory.

We evaluated the electron transmission  through
one ring or a series of rings as a function of the SOI strengths $\alpha$,
$\beta$,  the orientations
$\gamma_1$ and  $\gamma_2$ of the corresponding fields,  and the orientation $\gamma_3$
of the magnetic field for various parameters. As all figures demonstrate, the  transmission shows
a rich nontrivial structure with well-pronounced gaps as a function of any of these variables. For
a  series of rings these gaps acquire a square-wave shape, cf. Fig. 6, similar to that reported
for $\beta=0$ and $\gamma_1=\gamma_3=0$ \cite{mol1}. We also studied the case with the SOI present only in one arm of the ring(s) and saw how the oscillation pattern in the transmission changes due to the changes in the AC phase. If we change the   parameters from one ring to another in a series of rings, we can further modulate the transmission versus the SOI strength,
cf. Fig. 7 where the radius is changed.

A particular case of interest is that of the transmission, at zero magnetic field, as a function of the SOI strength
when the incident energy is such that $ka$ is a half integer. As elaborated at the end of  Sec. III B and shown in Fig. 8, the transmission is identical for any {\it odd} number of rings. If the number of rings is {\it even},  the transmission vanishes at the same points as that
for one ring and otherwise takes the highest constant value.  A simple explanation holds for $\beta=0$  and $\gamma_1=0$ \cite{mol1}. We confirmed numerically that this holds in our much more complicated case.  As shown though in Fig. 8, this breaks down when a small magnetic field is present.

For an incident electron initially spin-oriented along
the direction of propagation ($x$), the influence on the transmission of the RSOI
and DSOI terms, with effective electric field along the $z$ direction, is  identical
with that when the magnetic field is perpendicular  to the ring. Otherwise, the  RSOI
and DSOI affect the transmission in different ways.

For weak magnetic fields $B\leq 1$ T and realistic values of the SOI strength, the $g$ factor,
and the effective mass,  we showed the Zeeman term can be treated by perturbation theory and
has negligible effect on the transmission.

\section{Acknowledgement}
 This work was supported by the  Canadian NSERC Grant No.
OGP0121756.


\begin{references}
\bibitem{berr}M. V. Berry, Proc. R. Soc. London, Ser. A {\bf 392}, 45 (1984).

\bibitem{geom} {\it Geometric Phases in Physics}, edited by A. Shapere and F. Wilczeck (World Scientific, Singapore, 1989).

\bibitem{ahar1} Y. Aharonov and J. Anandan, Phys. Rev. Lett. {\bf 58}, 1593 (1987).

\bibitem{qian} T.-Z. Qian and Z.-B. Su, Phys. Rev. Lett. {\bf 72}, 3211 (1994).

\bibitem{sun} Q.-F. Sun, J. Wang, and H. Guo, Phys. Rev. B {\bf 71}, 165310 (2005).

\bibitem{ahar2} Y. Aharonov and D. Bohm, Phys. Rev. {\bf 115}, 485 (1959).

\bibitem{ahar3} Y. Aharonov and A. Casher, Phys. Rev. Lett. {\bf 53}, 319 (1984).

\bibitem{feve} P. Pfeffer, Phys. Rev. B 59, 15902 (1998); R. Winkler, {\it Spin-orbit coupling effects in two-dimensional electron and hole
systems}, Springer Tracts in Modern Phys. Vol. 191 (2003).

\bibitem{yi}Y.-S. Yi, T.-Z. Qian, and Z.-B. Su, Phys. Rev. B {\bf 55}, 10631 (1997);
M. Hentschel, H. Schomerus, D. Frustaglia, and K. Richter, Phys. Rev. B {\bf 69}, 155326 (2004).


\bibitem{gefe} Y. Gefen, Y. Imry, and M. Ya. Azbel, Phys. Rev. Lett. {\bf 52}, 129 (1984);
M. B\"{u}ttiker, Y. Imry, M. Ya. Azbel, Phys. Rev. A {\bf 30}, 1982 (1984).

\bibitem{aron}A. G. Aronov and Y. B. Lyanda-Geller, Phys. Rev. Lett. {\bf 70}, 343 (1993).

\bibitem{xia} J. B. Xia, Phys. Rev. B \textbf{45}, 3593 (1992); T. Choi, S. Y.
Cho, C. M. Ryu, and C. K. Kim, {\it ibid} \textbf{56}, 4825
(1997).

\bibitem{moln} B. Moln\'{a}r, F. M. Peeters, and P. Vasilopoulos, Phys. Rev. B \textbf{69}, 155335 (2004).

\bibitem{nitt}J. Nitta, F. E. Meijer, and H. Takayanagi, Appl. Phys. Lett. {\bf 75}, 695 (1999).

\bibitem{kog} T. Koga, Y. Sekine, and J. Nitta, cond-mat/0504743.

\bibitem{frus} D. Frustaglia and K. Richter, Phys. Rev. B {\bf 69}, 235310 (2004); S. Souma and B. K. Nikolic, {ibid} {\bf 70}, 195346 (2004).

\bibitem{loss} D. Loss, P. Goldbart, and A. V. Balatsky, Phys. Rev. Lett. {\bf 65}, 1655 (1990).

\bibitem{bych}  Y. A. Bychkov and E. I. Rashba, J. Phys. C {\bf 17}, 6039 (1984).

\bibitem{dres} G. Dresselhaus, Phys. Rev. {\bf 100}, 580 (1955).

\bibitem{ross} U. Rossler and J. Kainz, Solid State Communications {\bf 121}, 313 (2002).

\bibitem{sch} J. Schliemann, J. C. Egues, and D. Loss, Phys. Rev. Lett. {\bf 90}, 146801 (2003).

\bibitem{moro} A. V. Moroz and C. H. W. Barnes, Phys. Rev. B {\bf 60}, 14272 (1999).

\bibitem{rash} E. I. Rashba, Phys. Rev. B {\bf 70}, 201309(R)(2004); A. G. Mal'shukov and K. A. Chao, Phys. Rev. B 71, 121308(R) (2005).

\bibitem{meij} F. E. Meijer, A. F. Morpurgo, and T. M. Klapwijk, Phys. Rev. B {\bf 66}, 33107 (2002); Y. C. Zhou, H. Z. Li, and X. Xue, Phys. Rev. B {\bf 49}, 14010 (1994).

\bibitem{grif} S. Griffith, Trans. Faraday Soc. \textbf{49}, 345 (1953).

\bibitem{mol1} B. Moln\'{a}r,  P. Vasilopoulos,  and F. M. Peeters, Appl. Phys. Lett. \textbf{85}, 612 (2004).




\end{references}
\end{document}